\documentclass[floatfix,prd,twocolumn,superscriptaddress,showpacs,amssymb,amsmath,amsfonts,aps,altaffilletter,nofootinbib,nopreprintnumbers,showpacs,longbibliography]{revtex4-1}
\usepackage[latin9]{inputenc}
\usepackage{float}
\usepackage{graphicx}
\usepackage{esint}
\usepackage[colorlinks]{hyperref}

\makeatletter

\@ifundefined{textcolor}{}
{%
 \definecolor{BLACK}{gray}{0}
 \definecolor{WHITE}{gray}{1}
 \definecolor{RED}{rgb}{1,0,0}
 \definecolor{GREEN}{rgb}{0,1,0}
 \definecolor{BLUE}{rgb}{0,0,1}
 \definecolor{CYAN}{cmyk}{1,0,0,0}
 \definecolor{MAGENTA}{cmyk}{0,1,0,0}
 \definecolor{YELLOW}{cmyk}{0,0,1,0}
}

\makeatother

\begin{document}

\title{Reducing the number of templates for aligned-spin compact binary coalescence gravitational wave searches using metric-agnostic template nudging}

\author{Nathaniel Indik}
\affiliation{Max-Planck-Institut f\"ur Gravitationsphysik (Albert-Einstein-Institut), Callinstr. 38, D-30167 Hannover, Germany}
\affiliation{Leibniz Universit\"at Hannover, Welfengarten 1-A, D-30167 Hannover, Germany}

\author{Henning Fehrmann}
\affiliation{Max-Planck-Institut f\"ur Gravitationsphysik (Albert-Einstein-Institut), Callinstr. 38, D-30167 Hannover, Germany}
\affiliation{Leibniz Universit\"at Hannover, Welfengarten 1-A, D-30167 Hannover, Germany}

\author{Franz Harke}
\affiliation{Max-Planck-Institut f\"ur Gravitationsphysik (Albert-Einstein-Institut), Callinstr. 38, D-30167 Hannover, Germany}
\affiliation{Leibniz Universit\"at Hannover, Welfengarten 1-A, D-30167 Hannover, Germany}

\author{Badri Krishnan}
\affiliation{Max-Planck-Institut f\"ur Gravitationsphysik (Albert-Einstein-Institut), Callinstr. 38, D-30167 Hannover, Germany}
\affiliation{Leibniz Universit\"at Hannover, Welfengarten 1-A, D-30167 Hannover, Germany}

\author{Alex B. Nielsen}
\affiliation{Max-Planck-Institut f\"ur Gravitationsphysik (Albert-Einstein-Institut), Callinstr. 38, D-30167 Hannover, Germany}
\affiliation{Leibniz Universit\"at Hannover, Welfengarten 1-A, D-30167 Hannover, Germany}

\date{\today}
\begin{abstract}

  Efficient multi-dimensional template placement is crucial in
  computationally intensive matched-filtering searches for \emph{Gravitational Waves} (GWs). Here,
  we implement the \emph{Neighboring Cell Algorithm} (NCA) to improve
  the detection volume of an existing
  \emph{Compact Binary Coalescence} (CBC) template bank. This algorithm has
  already been successfully applied for a binary millisecond pulsar
  search in data from the Fermi satellite. It
  repositions templates from over-dense regions to under-dense regions and
  reduces the number of templates that would have been required by a stochastic method to
  achieve the same detection volume. Our method is readily
  generalizable to other CBC parameter spaces. Here we apply
  this method to the aligned--single-spin neutron-star--black-hole
  binary coalescence inspiral-merger-ringdown gravitational wave
  parameter space. We show that the template nudging algorithm
  can attain the equivalent effectualness of the stochastic method with 12\%
  fewer templates.
    
\end{abstract}
\maketitle

\section{Introduction}
\label{sec:Introduction}

Compact binary coalescence systems consisting of neutron-stars and/or black-holes
are key targets for the present generation of gravitational wave
detectors such as Advanced LIGO \cite{aligo} and Advanced Virgo
\cite{avirgo}.  The LIGO and Virgo detectors have observed a number of
binary black-hole coalescence events  \cite{Abbott:2016blz, Abbott:2016nmj,Abbott:2017vtc,Abbott:2017gyy,Abbott:2017oio} and recently a binary neutron-star coalescence event \cite{TheLIGOScientific:2017qsa}. The detection of
a coalescence of a neutron-star--black-hole binary has yet to be seen, but is increasingly likely
to be observed in the future. The
searches for compact binaries \emph{a priori} cover a wide range of masses
and spin magnitudes. A key ingredient in these searches is a suitable
template bank, i.e. a collection of model gravitational wave signals
which cover the desired parameter space.  

For searches based on matched filtering with modeled waveforms, the
traditional method of constructing a template bank was to use the
parameter space metric \cite{Owen:1998dk,Dhurandhar:1992mw} for
determining the spacing between adjacent templates.  This method was successfully demonstrated for non-spinning systems
\cite{Cokelaer:2007kx,Abbott:2009tt} and aligned-spin
systems \cite{Brown:2012qf,Harry:2013tca}.  In situations where the
parameter space structure is not sufficiently well
understood, stochastic methods are used
\cite{Messenger:2008ta,Babak:2008rb,Harry:2009ea}.

Stochastic methods
place templates at random points in the parameter space which are
drawn from an initially chosen distribution.  The chosen template is
then compared with previously accepted templates, and accepted only
if it is sufficiently far away each of the previously accepted
templates.  The procedure terminates when a certain \emph{coverage} (the
ratio of rejected templates over the total number of template
candidates) has been achieved resulting in a final \emph{saturated} template bank.
These stochastic methods are more
generally applicable but they are typically less efficient than the
geometric methods, i.e. they require more templates than a geometric
bank to achieve the same effectualness over the same parameter space.
Furthermore, the construction of a stochastic template bank can be
computationally demanding since, in principle, each new proposed
template needs to be compared with previously chosen templates.
This problem becomes particularly acute the closer the
bank gets to saturation.  The computational problem also becomes especially demanding
when precession effects are considered as these additional degrees of freedom require a large number of templates 
\cite{Indik:2016qky,Harry:2016ijz}.  It is therefore important to
consider methods of optimizing a template bank, specifically finding ways of
improving effectualness for a given number of templates and reducing the
computational cost.

In this paper we shall meet this computational challenge and show how
stochastic methods can be improved by employing the nearest neighbor cell algorithm (NCA)
\cite{Fehrmann:2014cpa, Pletsch:2012sh} and repositioning templates to
maximize the \emph{effectualness} and \emph{detection volume}.  It was
shown in \cite{Fehrmann:2014cpa} that for the Fermi $\gamma$-ray
pulsar search, using these methods leads to a reduction in the number
of distance computations in three dimensions by about five orders of
magnitude compared to other standard stochastic template bank
algorithms.  Here we shall apply these ideas to the Gravitational Wave (GW) Compact Binary Coalescence (CBC)
problem. Specifically, we shall focus on neutron-star--black-hole
(NSBH) systems since they make up $60\%$ of the templates placed in the last LIGO-Virgo
observation period \cite{TheLIGOScientific:2016qqj,DalCanton:2017ala}. We expect that our method would apply to other
source systems as well.  We consider NSBH binaries with a black-hole of mass $M_{BH}$ and a neutron-star mass of $M_{NS}$ such that
$2M_\odot < M_{BH} < 16M_\odot$, and $1M_\odot < M_{NS} < 3M_\odot$.  We use the inspiral-merger-ringdown-phenomenological waveform model
(IMRPhenomD) \cite{Santamaria:2010yb,Hannam:2013oca,Khan:2015jqa} to approximate the underlying
coalescence NSBH gravitational wave signal.

The plan for the paper is as follows.
Sec.~\ref{sec:background} briefly summarizes the necessary background
material and introduces notation that will be used later. In
particular we defines the \emph{maximal mismatch}, effectualness and detection
volume associated with a template bank.  It also describes the
\emph{template nudging algorithm}.  Sec.~\ref{sec:chirptime} applies this to
NSBH aligned spin templates.  It first introduces the chirp time
coordinates and calculate the template
isosurfaces at some selected points in parameter space
in the absence of a metric (as opposed to the method in \cite{Fehrmann:2014cpa, Pletsch:2012sh}
which requires an analytic expression for the metric).
Sec.~\ref{sec:NSBHbank} applies the optimization scheme to the
aligned spin template bank and finally Sec.~\ref{sec:results} presents
our results which are followed by concluding remarks.

\section{Background}
\label{sec:background}

\subsection{Matched filtering}
\label{sec:Matched Filtering}

Matched filtering is a methodology used to determine if time series
data $x(t)$ (where $t$ denotes
time), contains some signal with parameters $p_I$ of known form,
$h(p_T | t)$, or only instrumental noise $n(t)$. Thus, in the absence
of a signal,
\begin{equation}
  x(t) = n(t) \,,
\end{equation}
and in the presence of a signal
\begin{equation}
  x(t) = h(p_I | t) + n(t) \,.
\end{equation}
If the noise is stationary, we can characterize it by the single-sided
power-spectral-density (PSD) $S_n(f)$ according to
\begin{equation}
  \langle \tilde{n}^\star(f) \tilde{n}(f^\prime)\rangle \equiv \frac{1}{2} S_n(f)\delta(f-f^\prime)\,.
\end{equation}
Here the brackets $\langle\cdot\rangle$ denote an average expectation value over many
realizations of the noise, and $\tilde{n}(f)$ denotes the Fourier
transform of $n(t)$.

The PSD is used to define the inner product between two time-series $x(t)$ and $y(t)$:
\begin{equation}
  (x|y) \equiv 4 \textrm{Re}\int_0^\infty \frac{\tilde{x}^\star(f)\tilde{y}(f)}{S_n(f)} df\,.
\end{equation}
This inner product is used to define the norm of a time series $x(t)$
and a normalized time series $\hat{x}$ in the usual way: 
\begin{equation}
  ||x|| \equiv (x|x)^{1/2} \,,\qquad \hat{x} = x/||x||\,.
\end{equation}
For Gaussian noise, the likelihood function $\Lambda$ is
\cite{Finn:1992wt,Jaranowski:1998qm}
\begin{equation}
  \log\Lambda \equiv (x(t)|h(p_I|t)) - \frac{1}{2}(h(p_I)|h(p_I))\,.
\end{equation}
The idealized procedure to search for a signal with unknown parameters
is to compute $\log\Lambda$ for all points (suitably
discretized) in a given parameter space and to find the point where $\log\Lambda$ is maximum.
The likelihood can be analytically maximized for certain parameters (such as the
initial phase $\phi_0$ and an overall constant amplitude) or by a Fast-Fourier
transform (such as the time of arrival $t_0$) (see
e.g. \cite{Allen:2005fk}), while other parameters must be explicitly maximized over. These parameters we denote as
$\lambda_i$.

A template bank is a collection of waveforms $\{h(p_T^j)\}$ with parameters $\{p_T^j\}$, labeled by the index
$i$.
Given a template bank, we would like to know how effective it is in
recovering a given signal $h_I$. This is quantified in terms of a number, namely the
\emph{fitting-factor} (FF) defined as,
\begin{equation}
  FF(h(p_I), \{h(p_T^j)\}) \equiv \max_{j} \mu(h(p_T^j), h(p_I))\,,
\end{equation}
where
 \begin{equation}
  \mu(h(p_T^j), h(p_I)) \equiv \max_{t_0,\phi_0} (h(p_T^j)|h(p_I)(t_0,\phi_0))
 \end{equation}
is the \emph{match} between $h(p_T^j)$ and $h(p_I)$. $\mu(h(p_T^j),h(p_I))$ represents the fraction
of the optimal signal-to-noise ratio (SNR) of signal $h(p_T^j)$ captured by the template $h(p_I)$. As matter of notational convenience, we define the \emph{mismatch} as 
 \begin{equation}
 mm(h(p_T^j),h(p_I)) \equiv 1 - \mu(h(p_T^j),h(p_I))
 \end{equation}
which represents the fraction
of the optimal SNR of signal $h(p_I)$ not captured by the template $h(p_T^j)$.
The fitting
factor depends on a particular template bank and
a particular signal $h_I$.  Since we will compute this
for a fixed template bank, for notational convenience we  usually drop its dependence on
$\{h(p_I)\}$ and write $FF(h_I)$.

The loss in SNR can be quantified by the match between a signal and the 
nearest template and can be formulated geometrically \cite{Owen:1995tm,Owen:1998dk}.
The match between nearby points in parameter space can be approximated as
\begin{equation}
  \mu(\hat h(\lambda), \hat h(\lambda+d \lambda)) = 1 - g_{ij}d\lambda^i d\lambda^j  + \ldots
\end{equation}
with the metric
\begin{equation}
g_{ij} = -\frac{1}{2} \left.\frac{\partial^2 \mu\left(\hat h(\lambda), \hat h(\lambda')\right)}{\partial \lambda'_i \partial \lambda'_j}\right|_{\lambda'= \lambda}\,.
\end{equation}
This metric\footnote{For Gaussian stationary noise, one can show that
  the metric $g_{ij}$ is equivalent to constructing the scalar product
  $\frac{1}{2}\left.\left( \frac{\partial \hat h}{\partial \lambda_i}
    \right|\frac{\partial \hat h}{\partial \lambda_j}\right)$
  and projecting out the parameters $t_0$ and $\phi_0$.}  is useful in
quantifying the density of templates. The higher the metric determinant,
the higher the required template density is for a given allowed SNR
loss (which corresponds to a given \emph{maximal mismatch}). In the case of the
parameter space considered in Section~\ref{sec:isosurface}, we shall see
that the methods presented in \cite{Fehrmann:2014cpa, Pletsch:2012sh}
must be generalized to deal with the challenge of \emph{a priori} not having an explicit
expression for a metric.

For aligned waveforms, there is an analytic expression for the
inspiral-only metric for the NSBH parameter space
\cite{Brown:2012qf,Harry:2013tca}.  However
there is no known analytic expression for the metric of the inspiral-merger-ringdown (IMR) NSBH precessing
parameter space \cite{OShaughnessy:2015xjs} . Numerical approximations to the
metric can be ill-conditioned in curved regions of the parameter
space.  Therefore, a robust method for repositioning, \emph{nudging}, templates must
depend on numerical mismatch calculations.

\subsection{Template bank Effectualness}
\label{sec:effectualness}

An optimally placed template bank minimizes the number of templates for a given volume and maximizes the \emph{detection volume} for a given prior distribution of sources. We consider the prior distribution of NSBH binaries selected from a uniform component mass and aligned BH spin distribution. This distribution allows us to quantify the \emph{effectualness} of recovering a range of NSBH injections independent of the underlying astrophysical distribution.

We quantify the \emph{effectualness} of the template bank by determining the minimum recovered fitting factors of $99.9\%$ of a population of an injection set. To quantify the relative improvement of two or more CBC template banks,
we calculate the relative improvement in \emph{detection
volume} \cite{Canton:2014ena}.  The detection volume, $\mathcal{V}$, is assumed to be proportional to the sum of the
cube of the product of the optimal SNR of the injections, $\rho_{i}$,
with the fitting factor, $FF_{i}$, obtained from attempting to recover
a set of injected NSBH signals,
\begin{equation}
\mathcal{V\propto\mathrm{\sum_{i}(FF_{i}\rho_{i})^{3}}} \,.\label{eq:detectorVol_propto}
\end{equation}
By taking the ratio of the \emph{detection volumes} of two template banks,
$\mathcal{V_{\mathrm{2}}}$ vs $\mathcal{V_{\mathrm{1}}}$ , we obtain the \emph{relative detection volume} quantifying which template bank will perform better in a search.

\subsection{Template nudging algorithm}
\label{sec:optimization}
\begin{figure*}
\includegraphics[scale=0.6]{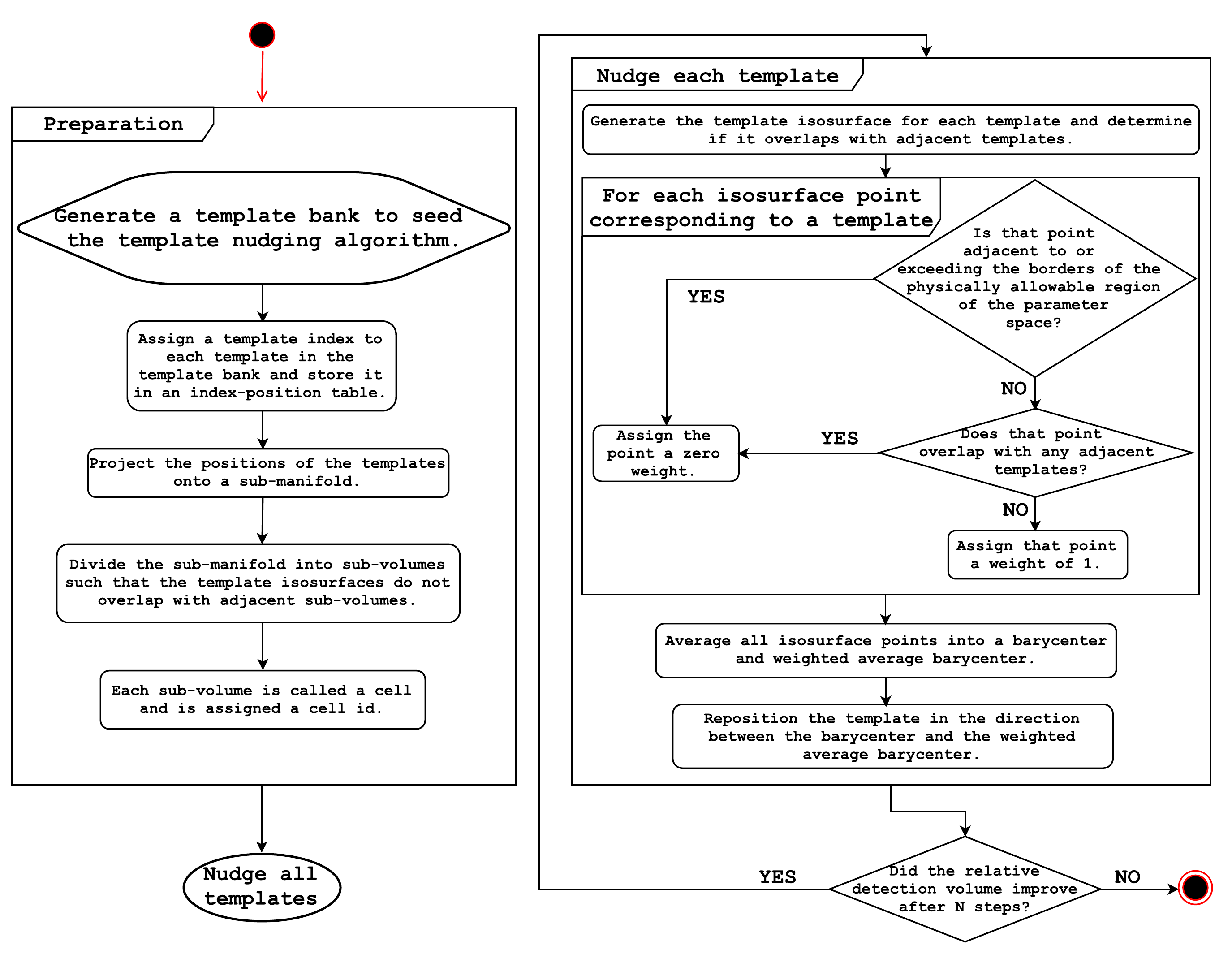}
\caption{Flowchart of the template nudging algorithm procedure. Preparation steps are
listed in the frame on the left and the template nudging procedure is listed in the frame on the right. In Section \ref{sec:results} we set ``N'' $=100$
for the termination condition shown in the diamond on the lower right.}
\label{fig:flowchart}
\end{figure*}
A stochastic placement algorithm is unlikely to place templates optimally
and will place more templates than are necessary to achieve the same effectualness 
\cite{Harry:2009ea}. We therefore employ an algorithm (see Figure \ref{fig:flowchart}) to move templates slightly, \emph{templates nudging},
facilitating the re-arrangement of templates in a template bank into
a configuration that improves the template bank's effectualness and detection volume 
without the addition of more templates.

In Section~\ref{sec:NSBHbank} we apply the template nudging algorithm 
to the CBC parameter space by adapting a version of the \emph{Neighboring Cell Algorithm} (NCA) method
\cite{Fehrmann:2014cpa} that does not require an analytic expression of the
metric and will work in parameter spaces where numeric approximations
of the generalized metric are ill-conditioned.
In our application of the method, the three dimensional regions covered by individual templates, template \emph{isosurfaces}, within a pre-specified maximal mismatch (i.e. 3\%) 
must be determined numerically.

The following list outlines the procedure for nudging a template bank in the absence of a metric.
\begin{enumerate}
	\item Select a template $T$.
         \item \label{surface_dist}
	  Find a set of points that are
	  uniformly distributed distributed on the boundary of $T$'s isosurface.
	\item \label{surface_check}
	  Check whether each of these points is
           inside a neighboring template's isosurface. If there is overlap, the considered
           boundary point gets zero \emph{weight}. If not, it gets unit weight (i.e. the template will not be nudged toward the adjacent overlapping template isosurface).
	 \item If a boundary point is outside of the considered
	   parameter space this point also gets zero weight.
	 \item  The boundary points are averaged together using these \emph{weights} into a \emph{barycenter}.
         \item  The template is nudged (i.e. coordinates are perturbed) in a direction determined by the barycenter offset relative to the unweighted barycenter of the boundary points and a maximum relative amount, $\epsilon$ (the pre-specified \emph{template nudge factor}, the fractional distance between the original template center and the closest isosurface point beyond which templates cannot be repositioned).
\end{enumerate}

This algorithm can be applied to any under-covered bank regardless of the method used to create that template bank (e.g. geometric, stochastic, or hybrid template bank placement methods which place templates with geometric and stochastic methods) and will improve the effectualness and detection volume. This provides a convenient way of enhancing any existing
template bank construction method.

\subsubsection{Neighboring cell algorithm}
\label{sec:Neighboring_cell_algorithm}

In order to efficiently cover the parameter space with non-overlapping cells,
the cells should be approximately the same size as the area of the chirp time volume covered by an individual template, as determined by the desired maximal mismatch between neighboring templates ($3\%$). 
In the simplest example these can be hyper spheres with a regular stacking in any choice of coordinates. To produce a list of the nearest neighboring templates we implement the following procedure.
	\begin{enumerate}
	\item Each cell is uniquely indexed.
	\item Given a cell index, the indices of neighboring
	  cells can be computed easily or can be stored in a
	  table.  Two cells are
	  neighboring if at least one point exists that has maximal mismatch
	  of $<3\%$ to each of the two cell regions.
	\item Template parameters are mapped to cell indices by a cell table. The parameters of each template lie
	  in the region of the corresponding cell.
	\item Given a position of a template, the index of a cell can
	  easily be computed by any kind of hash algorithm or a binary
	  search.  In
	  Euclidean space and with a hyper-cubic cell lattice this can
	  be achieved by using rounding or truncating operations on the
	  position values of the templates.
	\item A second table stores template indices and template
	  parameters in memory.
	\item Given a template, one finds all templates in the vicinity
	  by collecting the templates in the corresponding cell and all
	  neighboring cells. Therefore relatively few mismatches have to be
	  computed when placing a template as opposed to other 
	  stochastic placement algorithms.
\end{enumerate}

\section{Application to the CBC aligned-spin NSBH template bank}
\label{sec:NSBHbank}

\subsection{Aligned-spin binaries}
\label{sec:ASBinaries}

The parameter space of aligned-single-spin NSBH gravitational wave
signals considered in this paper can be represented with three
physical dimensions $\{M_{BH},M_{NS},\mathbf{S}_{BH}\cdot\hat{\mathbf{L}}\}$,
i.e. the masses of the black hole $M_{BH}$ and neutron star $M_{NS}$, and
the component of the black hole spin $\mathbf{S}_{BH}$ along the orbital
angular momentum $\mathbf{L}$ of the binary ($\hat{\mathbf{L}}$ is the
unit vector along $\mathbf{L}$).  We require further that
$-M^2_{BH}<\mathbf{S}_{BH}\cdot\hat{\mathbf{L}}<M^2_{BH}$ consistent
with a Kerr black hole in general relativity.  The BH spin representation
can also be expressed in a dimensionless form
\begin{equation}
  \mathbf{\chi}_{BH}=\frac{\mathbf{S}_{BH}\cdot\hat{\mathbf{L}}}{M^2_{BH}}\,.  
\end{equation}
We only consider binary systems with quasi-circular orbits and we
ignore any parameters associated with the internal
structure of the neutron-star.
We ignore neutron-star spins since they are expected
to be small, but the black-hole spin will be allowed to take any
magnitude which is meaningful in the Kerr metric and aligned/anti-aligned to the orbital angular momentum 
\cite{Lorimer:2008se}.\footnote{It is possible to approximate the
  effect of multiple spins by representing the spins as one \emph{effective spin} term. Gravitational waveforms like IMRPhenomD
  use {\it single-effective-one-body-spins} to model CBC binaries with
  two spinning component masses:
  $\{M_{BH},\mathbf{S}_{BH}\},\{M_{NS},\mathbf{S}_{NS}\}$. This effective
    spin is expressed as a single component mass weighted term,
  $\mathbf{S}_{EFF}=\left( M_{BH} \mathbf{S}_{BH} + M_{NS}
    \mathbf{S}_{NS}\right) / (M_{BH}+M_{NS})$.}
For computational simplicity, we transform these three physical
parameters into the Post Newtonian [PN] ``chirp time''
\cite{Poisson:1995ef} coordinates, $\{\tau_0,\tau_2,\tau_3\}$. In
these coordinates the underlying mismatch metric is approximately
flat at low frequencies.

\subsection{Isosurface geometry in chirp time coordinates} 
\label{sec:chirptime}
Template placement efficiency is largely dependent on the geometry of
the regions covered by individual templates, (i.e. the template
\emph{isosurfaces}). Using the physical parameters of the aligned-spin
NSBH parameter space $\{M_{BH},M_{NS},\chi_{BH}\}$, these template
isosurfaces are non-uniform.  An ideal coordinate system for
template placement would yield isosurfaces that are uniformly
spherical at any point in the proposal
distribution. Isosurfaces that have curved or sharp edges
are computationally challenging to model and tend to create
\emph{holes}, insufficiently populated regions in the template
bank.

Since we were unable \emph{a priori} to determine an ideal coordinate
system for the placement of NSBH IMRPhenomD templates, we
choose three chirp time coordinates $\{\tau_0,\tau_2,\tau_3\}$
\cite{Sathyaprakash:1991mt,Sengupta:2003wk} that have been demonstrated to flatten out
the TaylorF2 template bank \cite{Canton:2014ena,Brown:2012qf}. These
chirp times are defined with the following conventions where $f_0$ denotes a reference frequency, here chosen as $30$Hz:

\begin{eqnarray}
\tau_0 =&& \frac{5}{256} \frac{(\pi f_0)^{-\frac{8}{3}} (M_{BH}+M_{NS})^{\frac{1}{3}}}{M_{BH} M_{NS}}\\
\tau_2 =&& \frac{5}{256} \frac{M_{BH}+M_{NS}}{M_{BH} M_{NS}}(\pi f_0)^{-\frac{8}{3}} \nonumber \\
  && \times \left[\frac{743}{336}+\frac{11}{4} \frac{M_{BH} M_{NS}}{(M_{BH}+M_{NS})^2}\right]\\
\tau_3 =&& \frac{(\pi f_0)^{-\frac{5}{3}}}{128}(M_{BH}+M_{NS})^{\frac{32}{15}}
                         (M_{BH} M_{NS})^{-\frac{7}{5}}\nonumber\\ 
                         && \times
                          \Big{[}
                              16\pi - \frac{\chi_{BH}}{6} 
                              \left(
                                  \frac{19 M_{BH} M_{NS}}{(M_{BH}+M_{NS})^2}\right.\nonumber\\
                                  &&\quad\quad\quad\quad\quad\quad\quad\quad\left.+ \frac{113 M_{BH}}{M_{BH}+M_{NS}} +94
                               \right)
                           \Big{]}
\end{eqnarray}

\subsubsection{Template isosurfaces}
\label{sec:isosurface}

\begin{figure}
\includegraphics[scale=0.7]{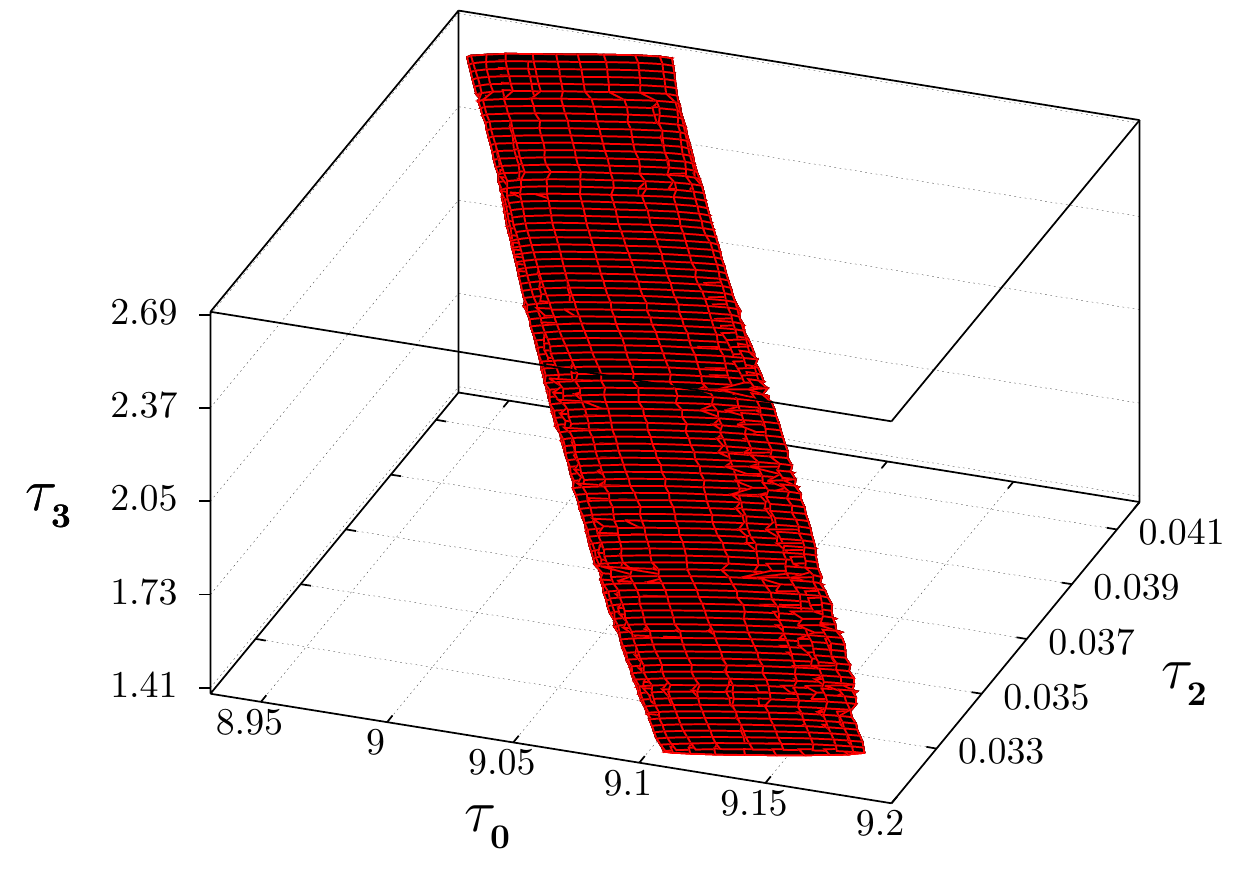}
\caption{Template isosurface corresponding to
  $\{M_{BH}=10M_\odot, M_{NS}=1.4 M_\odot, \chi_{BH}=0.5\}$ plotted in chirp
  time coordinates $\{\tau_0,\tau_2,\tau_3\}$. The isosurface
gets truncated as it hits the border of the physically allowable region of the parameter space.}
\label{fig:template}
\end{figure}

\begin{figure}

\includegraphics[scale=.7]{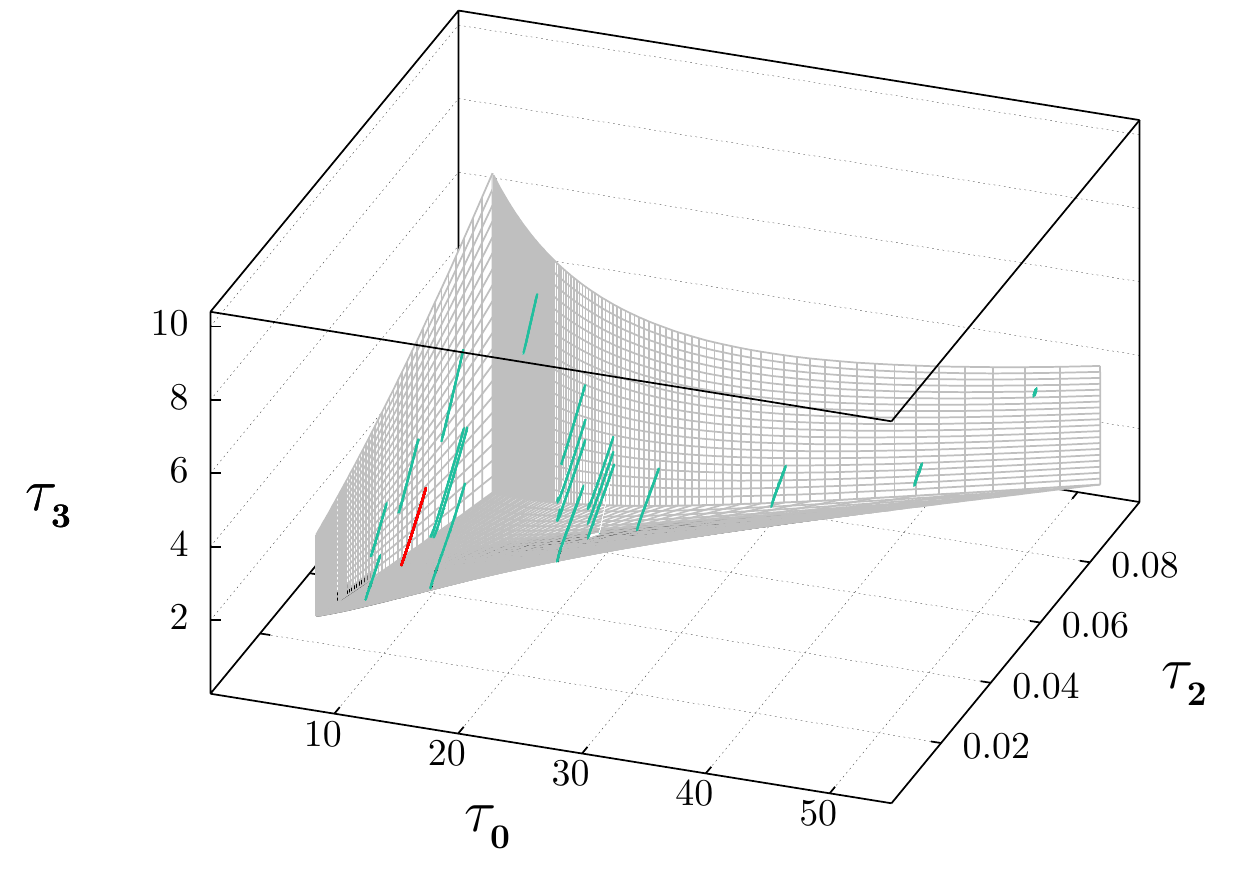}
\includegraphics[scale=.7]{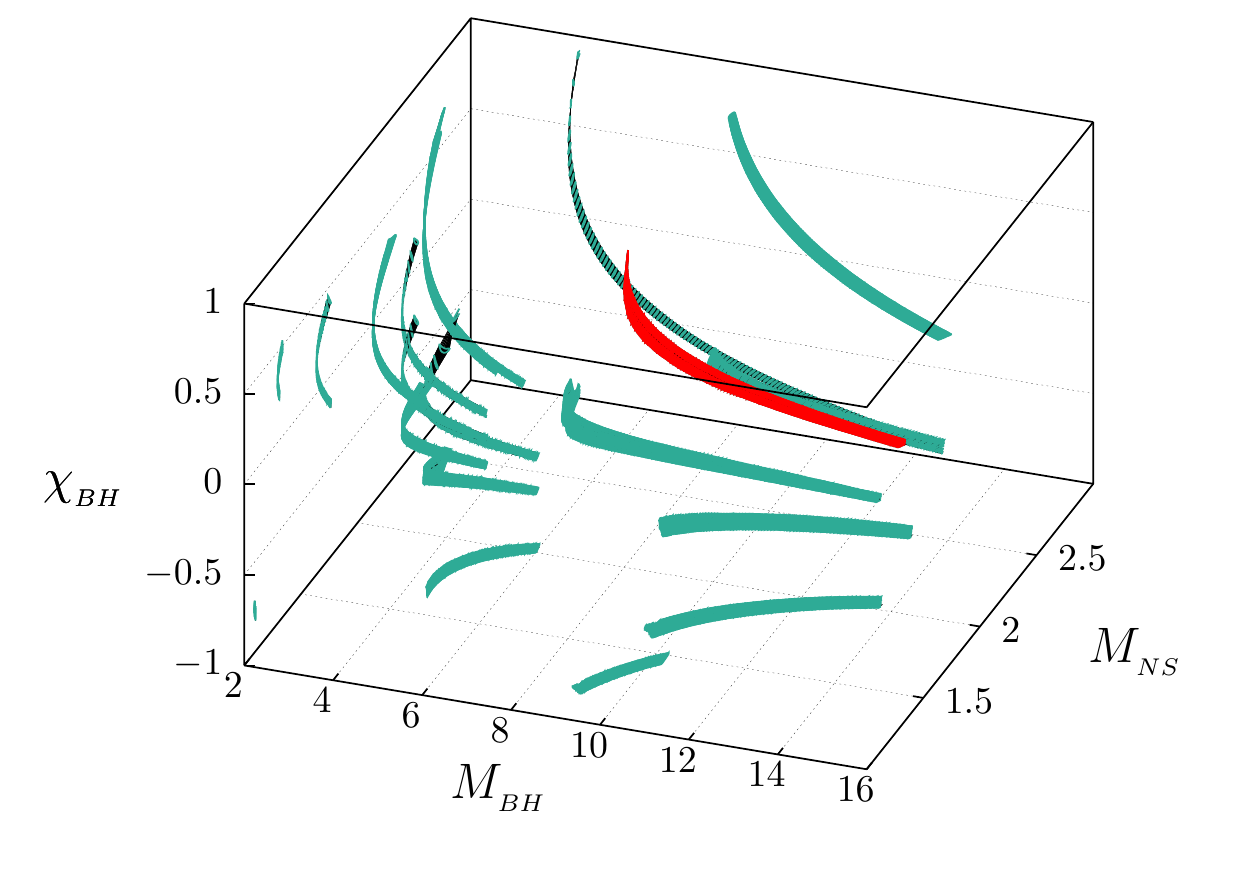}
\caption{Examples of template isosurfaces (dark green) within the borders of the physical allowable regions of the CBC single-spin NSBH parameter space. The top panel is plotted in chirp time coordinates $\{\tau_0,\tau_2,\tau_3\}$ (the gray mesh)  and the bottom panel in physical coordinates $\{M_{BH},M_{NS},\chi_{BH}\}$. The template isosurface corresponding to
  $\{M_{BH}=10M_\odot, M_{NS}=1.4 M_\odot, \chi_{BH}=0.5\}$ plotted in Figure~\ref{fig:template} is highlighted in red.}
\label{fig:templates}
\end{figure}

In these three chirp time coordinates, $\{\tau_0,\tau_2,\tau_3\}$, the regions of the
parameter space covered by individual templates are non-ellipsoidal, thin, 
and therefore difficult to model analytically (see Figure~\ref{fig:template} and ~\ref{fig:templates}). 

However, the IMRPhenomD waveform is least sensitive to perturbations in the $\tau_2$ degree of freedom. We found that individual cross-sections of these isosurfaces in the remaining two degrees of freedom can be modeled by two dimensional ellipsoids. We compensate for the lack of a three dimensional isosurface metric by modeling the three dimensional isosurface at a select number of cross-sections where the two dimensional projection of the isosurface is easier to model.

A further computational challenge is determining the scope of individual ellipsoid isosurface cross-section in the fewest computations possible. These ellipsoids are often very stretched out in these coordinates, therefore we implement the following method to select points.

The three dimensional mismatch isosurface can be obtained by computing
the mismatch isosurface of individual cross-sections by keeping 
$\tau_2$ constant for each cross-section. The isosurface is hence a ring
on the two dimensional sub-manifold parametrized by $\tau_0$ and
$\tau_3$. In the first step we find the point on this plane which has
the smallest mismatch with our considered template by applying a
simplex amoeba gradient-free-downhill method \cite{Nelder:1965zz}.
The routine starts by finding a point on the sub-manifold which is in the
allowed parameter space. The downhill method starts from there. Since
the template volume ranges in the $\tau_2$ direction from one end to the
other we always find a set of points on our considered $\tau_2$ plane which have
mismatches with our template smaller than the allowed critical
mismatch of $3\%$. These points are enveloped by the template isosurface and
are described by:
\begin{eqnarray}
  P &\equiv&
             \{\vec{p}_i | mm \left(\vec{p}_i,\vec{t}\right) \le 3\% \}\,,
\end{eqnarray}
where $\vec{t}$ is the vector of the template in the parameter space.
Usually we find $12$ to $25$ points by using the simplex amoeba gradient-free-downhill method. We use these ``inside'' points to obtain a local
approximation of the metric on this surface describing the
distance between the point of the maximal mismatch to the points
found by the simplex method. The description by a metric is not quite
correct since the mismatch of the minimal point is much smaller than $3\%$
but not zero. If this mismatch becomes significantly larger, then one might think about adding a constant to Eq.~\ref{MetricApproximation}. However, for deriving the isosurfaces the following method
works sufficiently well. We expand the metric starting with quadratic
terms:
\begin{eqnarray}
	\tilde{m}
&=&
	\sum_{p=2}^{\infty} \sum_{i_1 \ge i_2 \ge \cdots \ge i_p}^{D}
	\gamma_p (i_1,i_2,\cdots,i_p) \prod_{l=1}^{p} d_l
	\label{MetricApproximation}\,,
\end{eqnarray}
$D$ is the dimension of our manifold, in our case $D=2$ and $d_l$ are
the components of the distance vectors.
In the following we restrict ourselves to the second order expansion.  For
each point in the set, $P$ has a real mismatch $m_k$ and the
mismatch is approximated by the metric $\tilde{m}_k$. We choose the metric
components $\gamma_2 (i,j)$ such that the quantity $\chi^2 = \sum_k \left(m_k -
\tilde{m}_k\right)^2$ is minimized.

The component $d_k^i$ is the $i$th component of the $k$th distance vector towards our
minimal point $p_0$ on the plane.
We minimize the $\chi^2$ with respect to the metric components:
\begin{eqnarray}
	\frac{\partial \chi^2}{\partial \gamma_2(o,p)}
&=&
	\sum_k \left[
		m_k
		-
		\right.
		\\
&&
		\sum_{i_1 \ge i_2}^D
		\left.
		\gamma_2(i_1,i_2) d^k_{i_1} d^k_{i_2}
		\right] d^k_od^k_p
		\\
&=&
		0\,.
\end{eqnarray}
This set of equations can be described by
\begin{eqnarray}
&\Rightarrow&
	\underbrace{
	\sum_k
	\left(
	\begin{array}{ccc}
	  \mathcal{D}^k_{0} \mathcal{D}^k_{0} & \mathcal{D}^k_{0} \mathcal{D}^k_{1} & \mathcal{D}^k_{0} \mathcal{D}^k_{2} \\
	  \mathcal{D}^k_{1} \mathcal{D}^k_{0} & \mathcal{D}^k_{1} \mathcal{D}^k_{1} & \mathcal{D}^k_{1} \mathcal{D}^k_{2} \\
	  \mathcal{D}^k_{2} \mathcal{D}^k_{0} & \mathcal{D}^k_{2} \mathcal{D}^k_{1} & \mathcal{D}^k_{2} \mathcal{D}^k_{2} 
	\end{array}
  \right)
  }_{={\bf \mathcal{D} }}
  \left(
	\begin{array}{c}
	  \gamma_2(1,1)\\
	  \gamma_2(2,1)\\
	  \gamma_2(2,2)
	\end{array}
  \right) \\
&=&
	\sum_k m_k
	\left(
	\begin{array}{c}
		\mathcal{D}_0^k \\
		\mathcal{D}_1^k \\
		\mathcal{D}_2^k
	\end{array}
  \right)\,,
\end{eqnarray}
where $\mathcal{D}_i^k = (d^k_1)^{2-i} (d^k_2)^i$.
We invert $\bf \mathcal{D}$ and obtain the second order expansion values
for the metric $\gamma_2(i,j)$ . This procedure can be expanded to
arbitrarily high orders of the metric expansion.
We can now approximate the metric in a quadratic form:
\begin{eqnarray}
	g
&=&
	\left(
	\begin{array}{cc}
	  \gamma_2(1,1) & \frac{1}{2}\gamma_2(2,1) \\
		\frac{1}{2} \gamma_2(2,1) & \gamma_2(2,2)
	\end{array}
  \right)
  \label{}\,.
\end{eqnarray}
We compute eigenvalues $v_i$ and eigenvectors $\vec{e}_i$ of this metric
and compute a set of $N$ points approximately in the vicinity of the isosurface.
$I = {\vec{p}_i}$ with
$\vec{p}_i = \sin(2 \pi i/N) \vec{e}_1 \sqrt{0.03/v_1}
+ \cos(2 \pi i/N) \vec{e}_2 \sqrt{0.03/v_2}$. These points are not
equidistant but sufficiently well distributed for our purposes.

We shift the points in the radial direction with respect to the center
point such that the points have a mismatch of exactly
$3\%$ using the Newton-Raphson method.

We compute the barycenter of our shifted set of points and repeat
the metric approximation method and the shifting to get an even better
set of points.

The scheme is applied for all distinct $\tau_2$ planes.

\subsubsection{Cell Structure}
\label{sec:cells}

In order to apply the NCA method, we provide an appropriate cell structure with
the following properties:
\begin{itemize}
  \item the cells are sufficiently small
  \item templates within a cell can reach the neighboring cells, but not
    next-nearest neighboring cells
  \item a cell index can be easily and quickly computed knowing the template
    parameter space points.
\end{itemize}
Since each template spans the entire parameter space in the $\tau_2$
direction, splitting the parameter space in this direction is not
possible. On the other hand, the templates have a very small size in
the $\tau_0$ direction so we can safely split the parameter space in
this direction. We split the parameter space in the
$\tau_0$ direction into $300$ slices. We mapped the $\tau_0$ coordinates into a
unit length parameter space $x= (\tau_0-4)/50$. The templates are curved
in the $\tau_3$ direction, thus further splitting is not directly feasible.
To get the templates in a compact form we applied an \emph{ad hoc} transformation $y =
((\tau_3/\tau_2-20 + 2.5/\tau_2)\tau_0-480)/2500$. The new coordinate
$y$ is normalized and ranges from
$0$ to $1$.
We split the parameter space into $40$ slices in this direction and obtained a
rectangular grid in the $x$-$y$ plane with $12,000$ cells.

We tested this
setup with a set of randomly distributed points. In the limit of
having only one cell we compute all mismatches smaller than $3\%$
correctly. If the cells become too small some of mismatches smaller
than $3\%$ will not be detected. This happens if the cells are so small
that overlapping templates are not in neighboring cells anymore, but,
for instance, in next-nearest neighboring cells.

\section{Results}
\label{sec:results}
\begin{figure}
\includegraphics[scale=0.45]{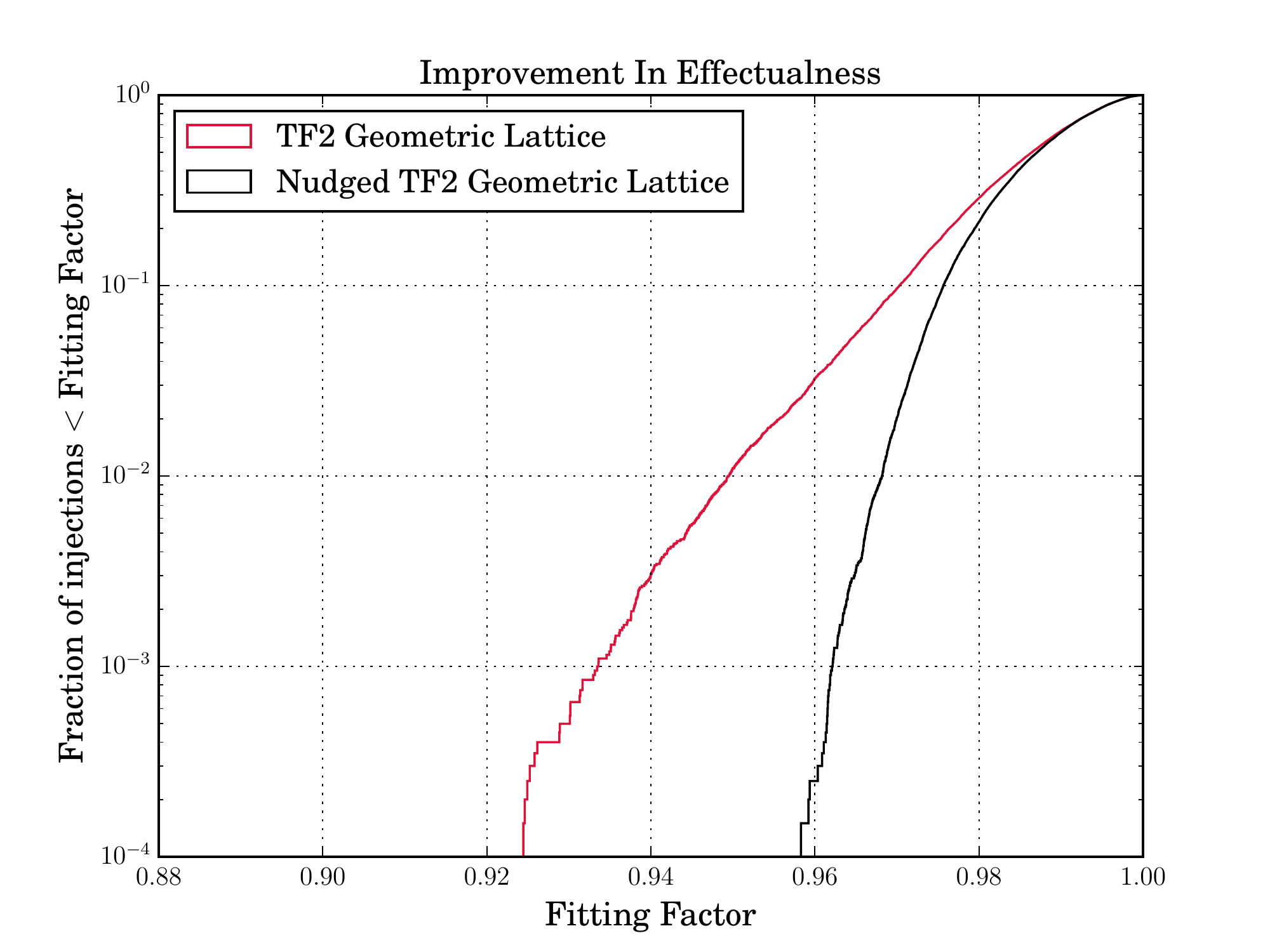}
\caption{The improvement in template bank effectualness
from applying the template nudging algorithm to a TaylorF2 (TF2) geometric lattice. This used $20,000$ IMRPhenomD injections 
drawn from a uniform component mass and aligned BH spin distribution. The maximum fitting factors of the lowest $0.1\%$ of injections improved from approximately $0.93$ to $0.96$.}
\label{fig:effectualness}
\end{figure}

To test the template nudging algorithm we seeded it with a geometric lattice TF2 template bank containing $174,000$ templates. Both the initial seed bank and the nudged bank were then tested against a set of $20,000$ random IMRPhenomD injections. For the nudged bank only $3\%$ of injections had a fitting factor less than $97\%$, compared to $10\%$ for the original seed geometric bank (see Figure~\ref{fig:effectualness}). 
For comparison, building a stochastic bank targeting $97\%$ \emph{minimal match} (i.e. $1-$maximal mismatch), required 220,000 templates, with only $0.1\%$ of templates having a fitting factor less than $97\%$ for the same injection set.

\begin{figure}
\includegraphics[scale=0.45]{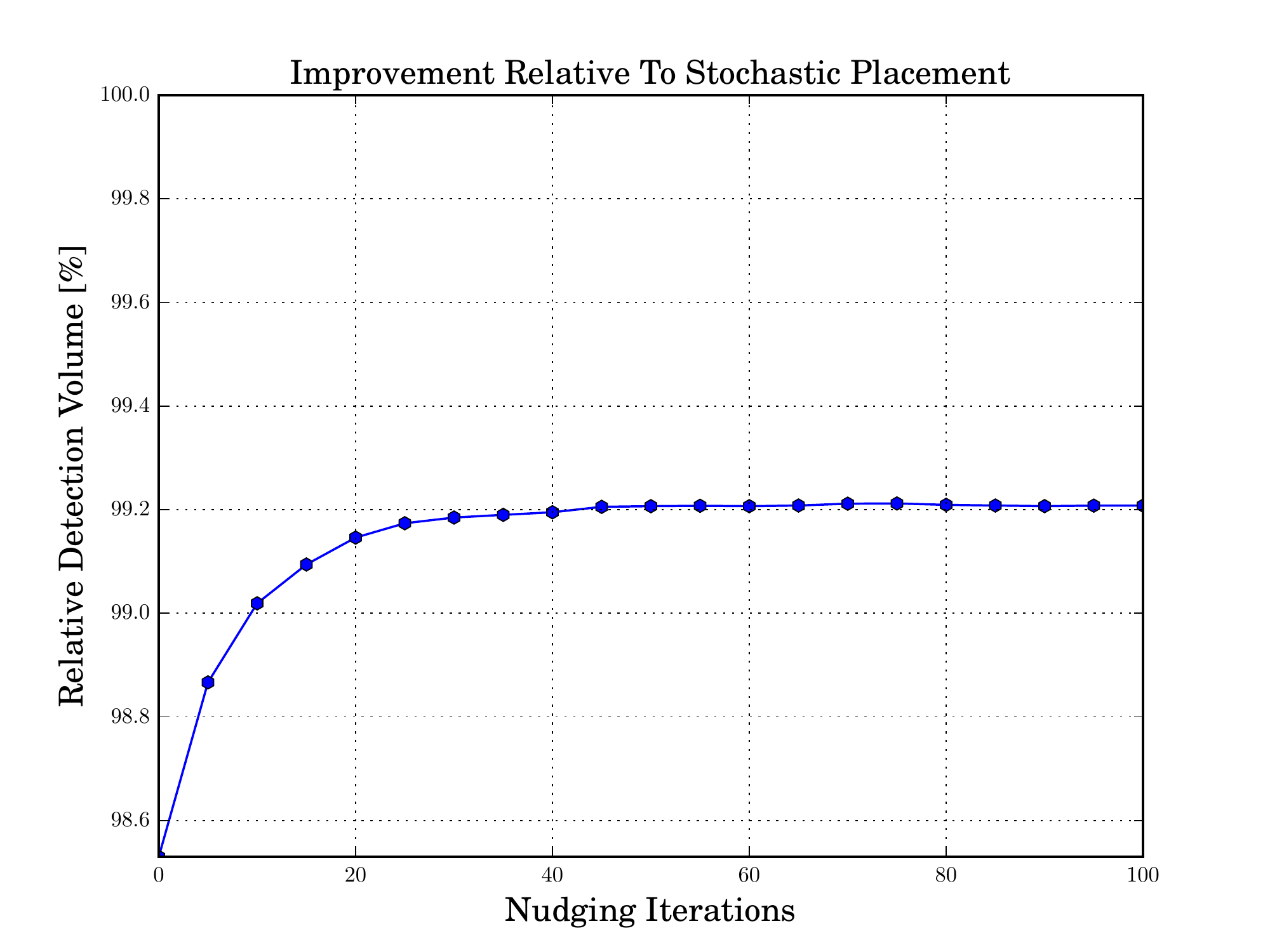}
\caption{The relative detection volume of the nudged bank versus the stochastic bank, $100\% * V_{nudged}/V_{stochastic}$, 
was tracked every five iterations of the template nudging algorithm using the same set 
of $20,000$ injections
drawn from a uniform component mass and aligned BH spin distribution. The improvement settles to a constant value after approximately $40$ iterations.}
\label{fig:sbank_convergence}
\end{figure}

\begin{figure}
\includegraphics[scale=0.45]{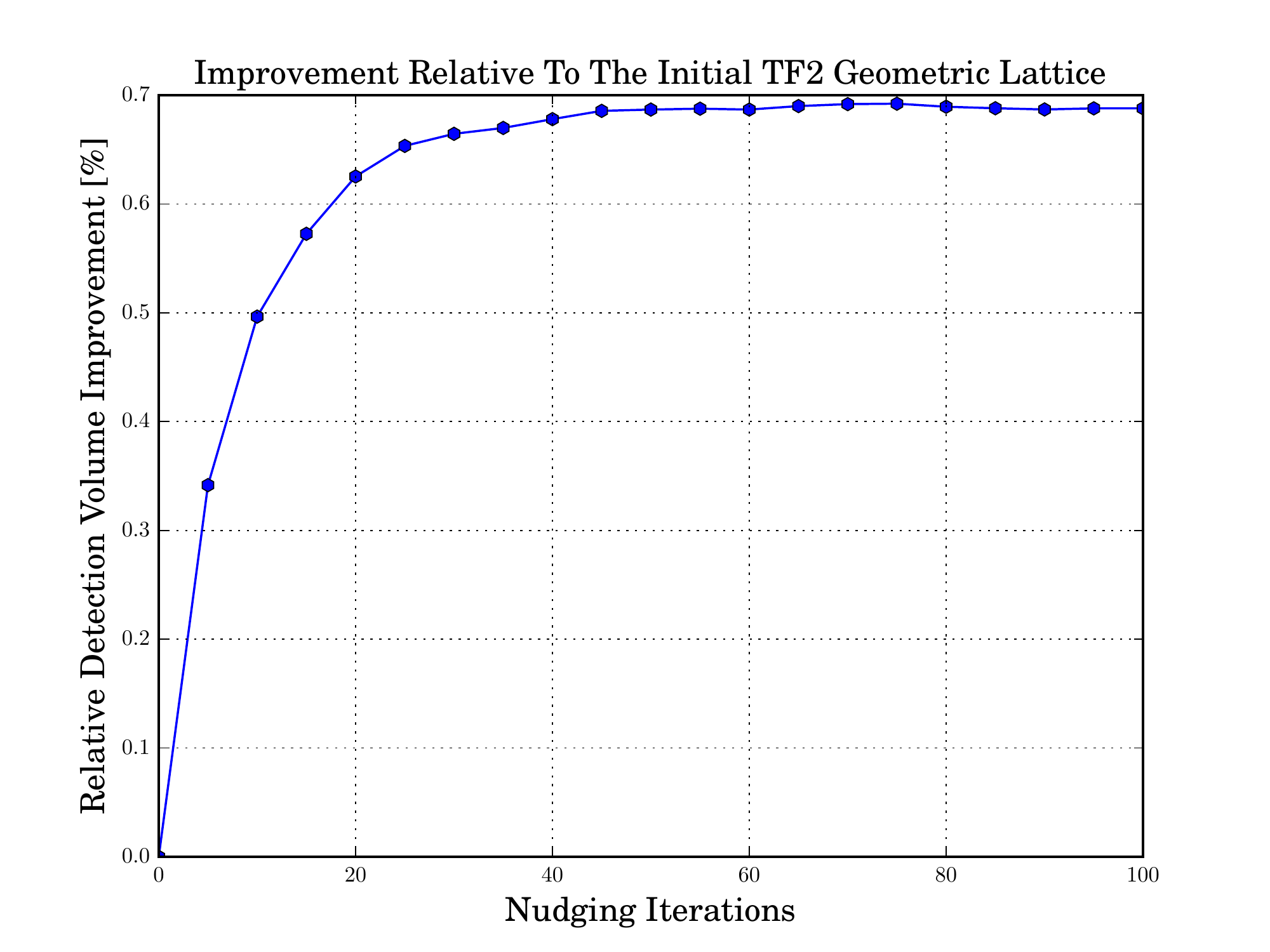}
\caption{The relative detection volume improvement of the nudged bank
versus the original TaylorF2 (TF2) geometric lattice bank,
 $100\% *( V_{nudged}/V_{original}-1)$,
was tracked every five iterations of the template nudging algorithm using the same $20,000$ injections
drawn from a uniform component mass and aligned BH spin distribution.}
\label{fig:geo_convergence}
\end{figure}

For this comparison matches were calculated between $30-1024$Hz using a PSD \cite{psd} built from the harmonic mean of the Hanford and Livingston PSDs taken within a few days of GW150914 and thus comparable to what shown in \cite{Abbott:2016blz}.
The stochastic bank was generated using \texttt{lalapps\_cbc\_sbank} \cite{pycbc,Ajith:2012mn,Capano:2016dsf,DalCanton:2017ala,LALsuite}, using
with a convergence criteria of rejecting $97\%$ trial templates. The stochastic algorithm is able to use IMRPhenomD templates directly to calculate matches.
The TaylorF2 lattice was generated using \texttt{pycbc\_geom\_aligned\_bank}
\cite{pycbc,Brown:2012qf,Capano:2016dsf,DalCanton:2017ala}, constructed with a two dimensional lattice such that each lattice point
had a maximal mismatch no larger than $3\%$. As can be seen in Figure~\ref{fig:effectualness}, this target was not attained for the IMRPhenomD test signals. A known metric is required to determine the geometric lattice. Since no metric is known for the IMRPhenomD templates we used TaylorF2 3.5PN waveforms \cite{Arun:2008kb,Bohe:2013cla,LALsuite} for which an analytical metric can be calculated. Although the lattice locations were calculated using TaylorF2 waveforms, these were subsequently swapped out with IMRPhenomD templates when testing the effectualness and this results in a loss of effectualness.

The template nudging algorithm was applied for $100$ iterations using one cross-section with at least $16$ surface points to approximate individual template isosurfaces for the nudging calculations. The effectualness and improvement in the relative detection volume between successive intermediate nudges was quantified by recovering a set of $20,000$ IMRPhenomD injections drawn from a uniform component mass and BH spin distribution (see Figures~\ref{fig:sbank_convergence} and~\ref{fig:geo_convergence}).

Within each iteration, templates were nudged in parallel for the purposes
of reducing computation time. Studying the effect of the template nudge factor on the convergence of the method revealed that a template nudge factor of $5\%$ produced the largest improvements in the template bank's effectualness in the first iteration relative to template nudge factor of $1\%$ or $.05\%$. However, in later iterations the variance of the improvement of the recovered fitting factors was also higher and a poorer fitting factor was recovered in the anti-aligned high mass regions of the bank relative to those produced by template nudge factor of $1\%$. We used a composite approach to improve the convergence of the method, wedding the advantages of using a bigger template nudge factor in the first iterations to the advantages of the greater precision of smaller template nudge factors in later iterations. Hence, we applied the template nudge algorithm in two batches of $50$ iterations per template nudge factors: $5\%$ and $1\%$.
After $100$ nudging iterations, the coverage had improved by
$3\%$ and the relative detection volume had increased by
$0.69\%$.

In order to extract an additional $1\%$ of
coverage and an additional $0.80\%$ relative detection
  volume beyond what is achieved with the nudging, we \emph{polished}
the nudged template bank by adding $20,000$ templates via a final stochastic placement to fill in any remaining holes. This produced a bank with overall fitting factors comparable to the stochastic bank, but with only $194,000$ templates compared to the $220,000$ templates required via the \texttt{lalapps\_cbc\_sbank} algorithm. Therefore it is possible to achieve equivalent template bank effectualness and detection volume with $26,000$ fewer templates than would be required by the purely stochastic method.

\begin{figure*}
\includegraphics[scale=0.435]{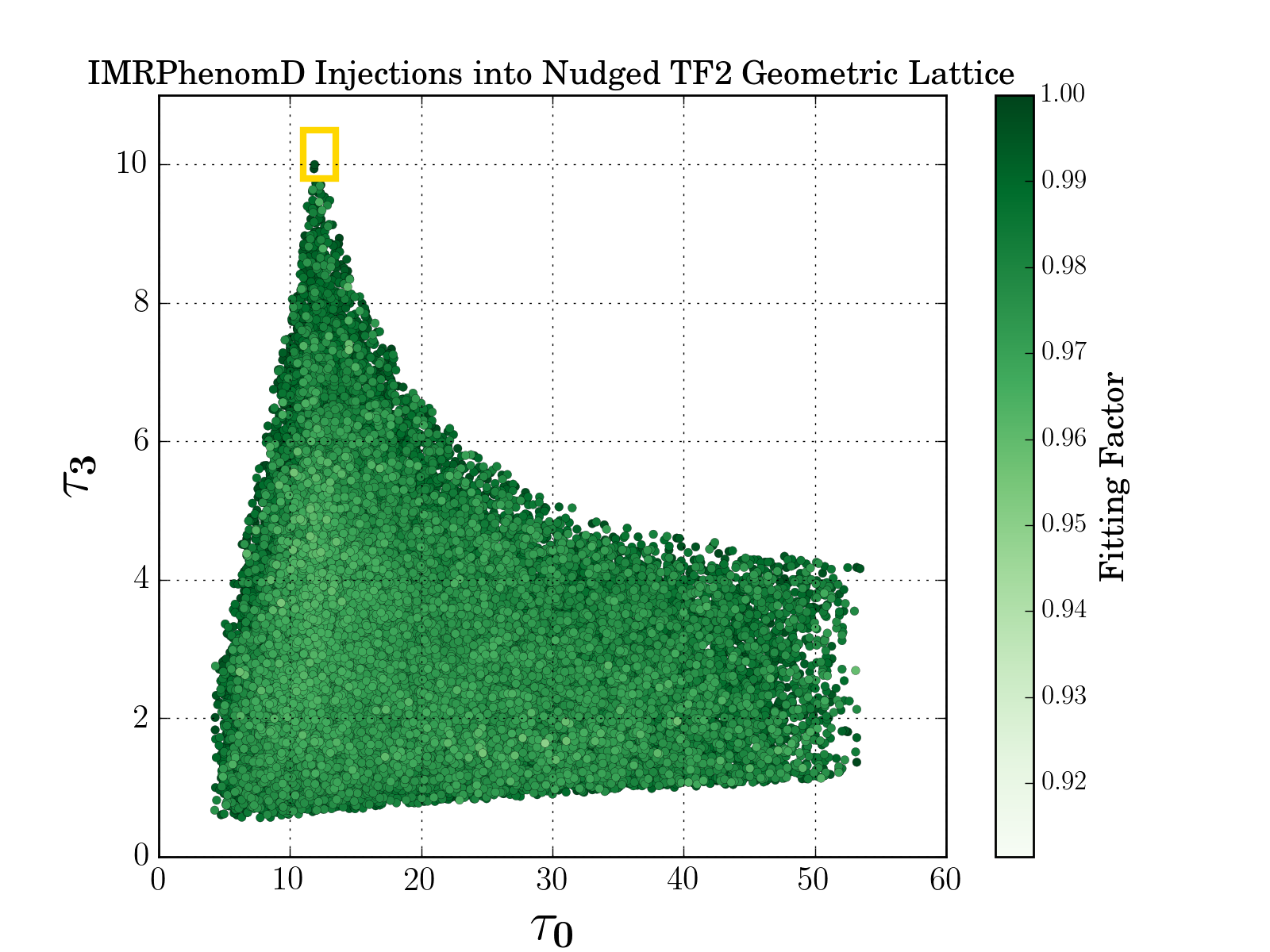}
\includegraphics[scale=0.435]{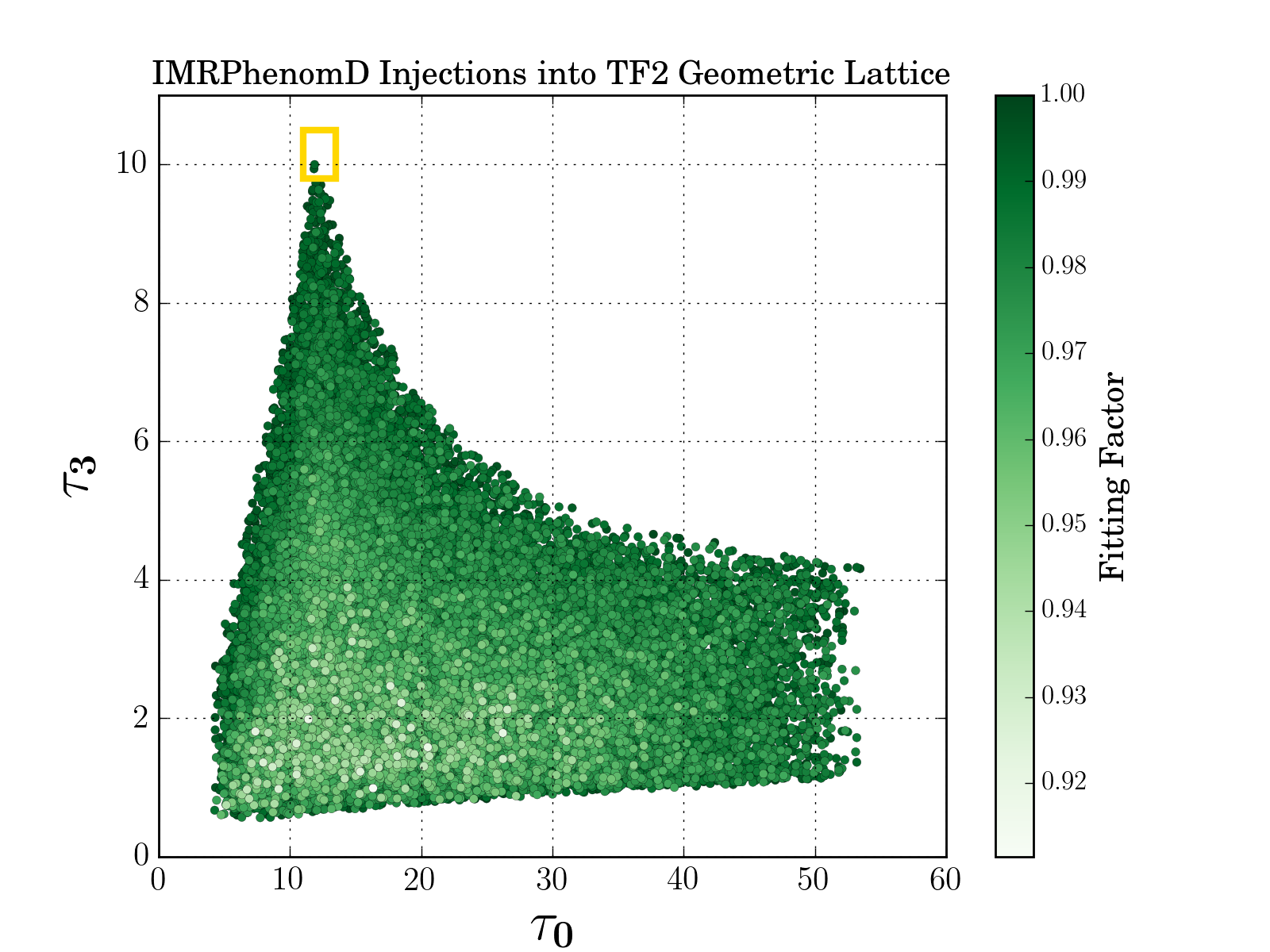}
\includegraphics[scale=0.435]{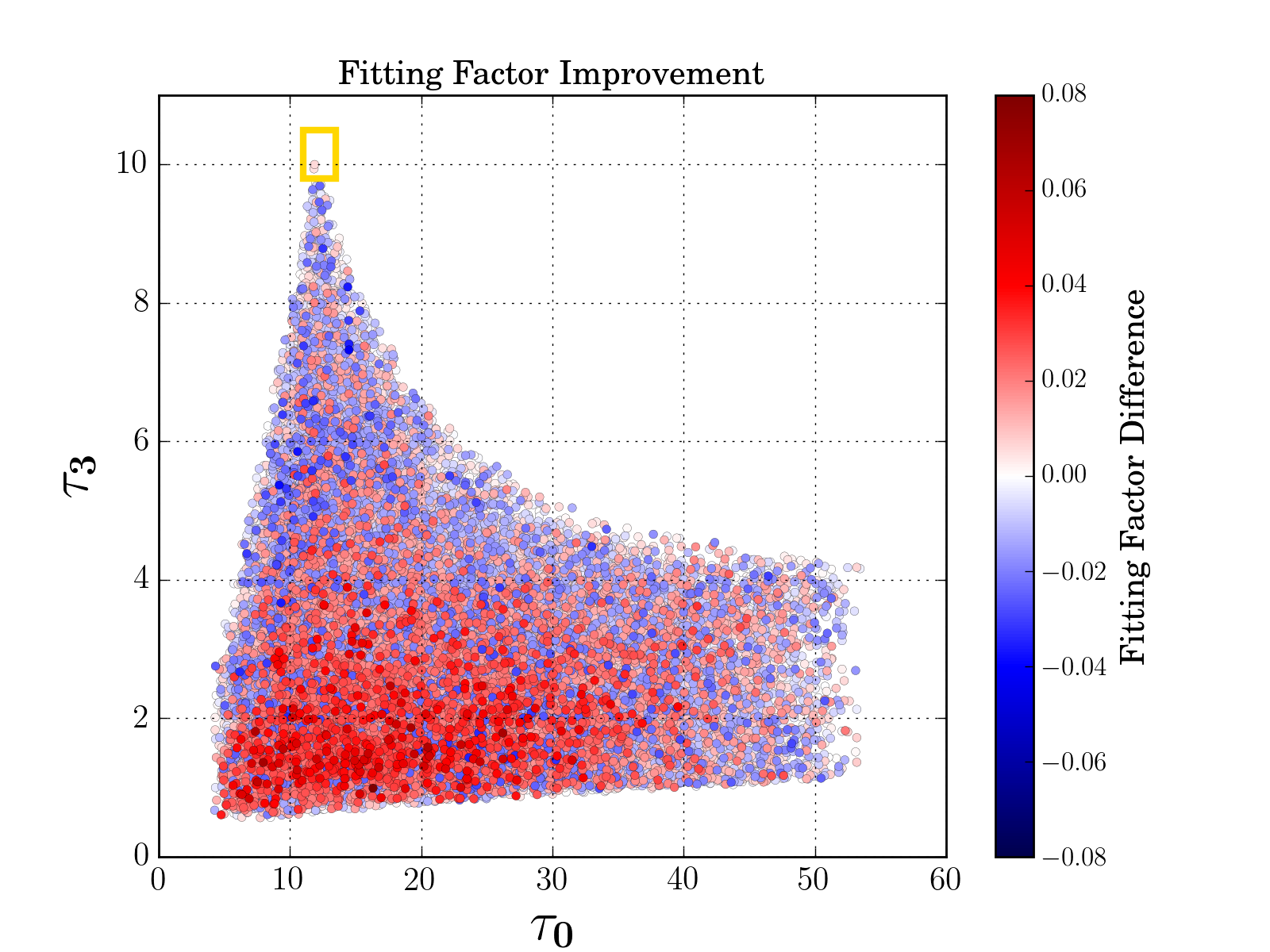}
\includegraphics[scale=0.435]{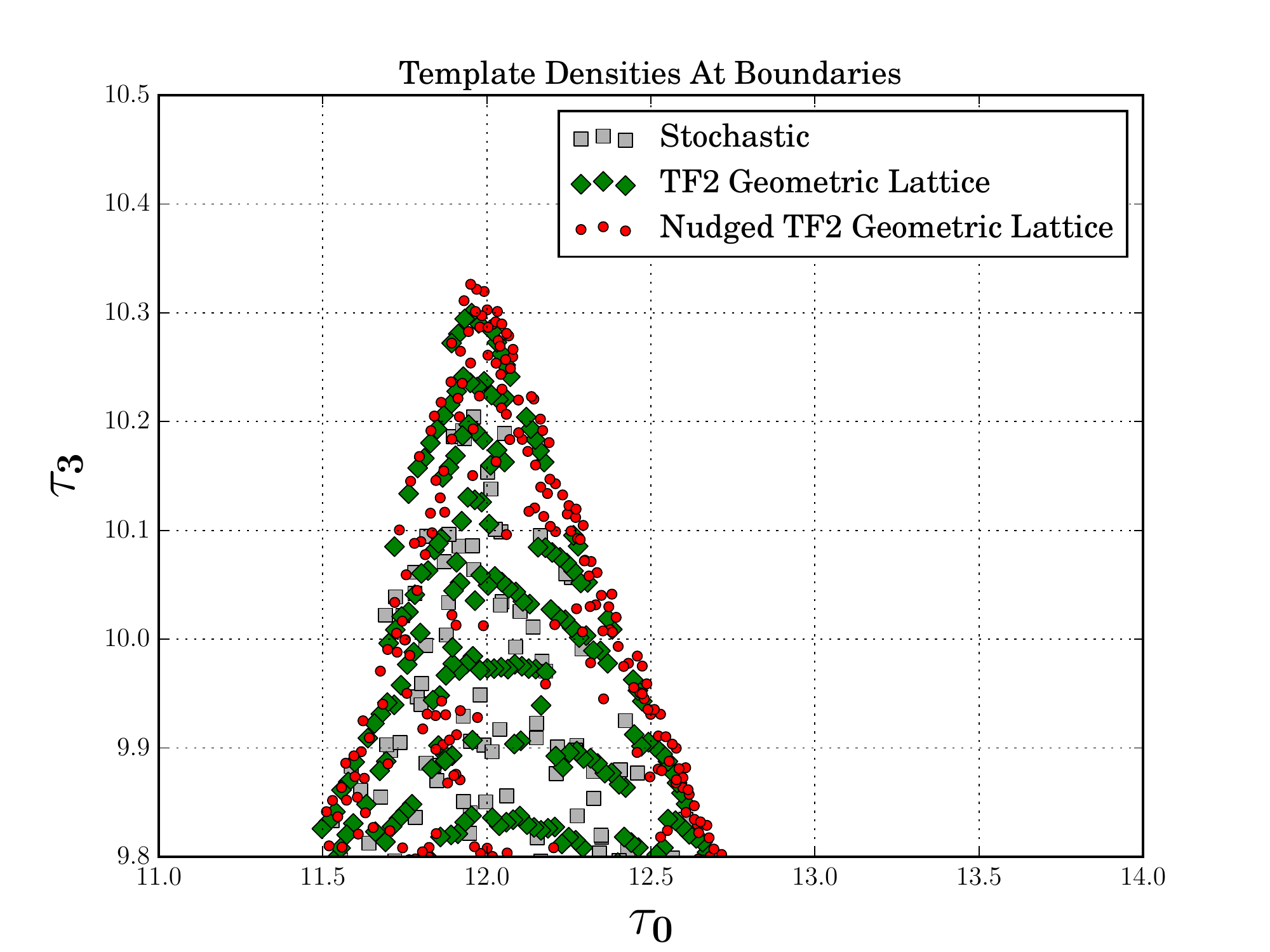}
\caption{Comparison of fitting factors for the template nudging algorithm. Coordinates are the chirp times given in Section~\ref{sec:chirptime}. The top row shows the  fitting factors for the nudged TF2 geometric lattice (left) and the original TF2 geometric lattice (right). The fitting factors are seen to be more uniform across the nudged bank. The difference in fitting factor is displayed in the bottom left panel. The regions of improvement (in red) are seen to correspond largely with the regions of insufficient fitting factor in the original geometric lattice. The lower right panel shows the template locations for three different template distributions in the highly anti-aligned spin region denoted by the yellow box in the other panels. In this region the nudging pushes templates to nearby boundaries and does not reduce the number of templates in a region that is over-covered by the geometric lattice.}
\label{fig:template_FF_comp}
\end{figure*}

Figure \ref{fig:template_FF_comp} compares the result of
injecting $40,000$ IMRPhenomD signals into the nudged TF2 geometric lattice 
and the original TF2 geometric lattice. The nudged TF2 geometric lattice has a more even distribution of
recovered fitting factors. The template nudging algorithm improves regions with poor fitting factors
by repositioning templates from over-covered regions. In principle this allows the attainment of a desired minimum fitting factor across the entire parameter space with fewer templates.
However, the template nudging algorithm as currently conceived does not remove templates from the bank. In some cases the nudging can result in templates from over-covered regions being nudged to nearby boundaries and piling up there. 
An example of this is shown in the lower right panel of Figure \ref{fig:template_FF_comp}. This highly anti-aligned spin region of the NSBH parameter space, indicated by the yellow boxes in the other panels, is over-covered by the original TF2 geometric lattice, obtaining recovered fitting factors close to $\sim 100\%$. In this region the excess templates are nudged to the boundaries and build up there, without reducing the fitting factors.

\section{Conclusions}
\label{sec:conclusions}
We have shown how the number of gravitational wave templates needed to search a region of parameter space can be reduced by repositioning templates. In particular, we successfully implemented a method to reduce the number of templates required by the algorithm \texttt{lalapps\_cbc\_sbank} to cover an NSBH single-aligned spin parameter space. This resulted in a $12\%$ reduction in the number of templates. For comparison, the hybrid method (utilized to build the O1 bank \cite{TheLIGOScientific:2016qqj} and the O2 bank \cite{DalCanton:2017ala}) required $5\%$ fewer templates than the stochastic method when used to build a binary-black-hole bank \cite{Capano:2016dsf}.
Given that $60\%$ of the templates in the O2 CBC template bank are in the NSBH mass range considered in this paper, applying the template nudging algorithm in this subspace alone would already reduce the size of the bank
placed on the entire mass range by $7\%$ without sacrificing effectualness. Assuming that the percentage reduction in the number of templates is uniform across the parameter space, then we could expect an overall reduction of approximately $50,000$ templates from the $400,000$ template O2 bank.

The template nudging is seen to be most effective at repositioning templates from over-covered regions to under-covered regions.
This is less effective in regions very close to boundaries of the desired parameter space.
In some cases the shifted templates can accumulate near the boundaries
as seen in the lower right panel of Figure \ref{fig:template_FF_comp}.
Once a template is nudged such that one or more of its
isosurface points exceeds the border of the bank,
there is currently no mechanism to nudge the template away from that border
or to remove it entirely.

For the nudging the template bank is split into two-dimensional planes and the nudging takes place within each plane. The algorithm does not nudge templates between the planes and thus if there is a hole in one of the planes the method is not able to fill it with excess templates from another plane. Such banks may require additional templates after nudging
in order to match the effectualness and detection volume
that can be produced by a stochastic method. In our example, this would occur specifically when these holes are in
regions of the bank which require nudging the $\tau_2$ coordinate since this version of the template nudging algorithm only
nudges templates on the $\tau_0-\tau_3$ plane. It is non-trivial to remove this restriction
since the template isosurface cross-sections in the $\tau_0-\tau_2$ and the $\tau_2-\tau_3$ planes are non-ellipsoidal and would require modifying how the algorithm samples the 
isosurface boundary points.

Further work is needed to adapt this method to the entire
aligned-spin CBC parameter space. Currently the algorithm only places
templates in the NSBH and Binary Neutron Star (BNS) mass range.
Extending this region would require readdressing the following two points: 1) defining the borders of the
targeted chirp-time parameter space and
2) obtaining a uniform transformation across the targeted chirp-time parameter space to the two dimensional grid needed to apply our modified NCA method.
In this paper, these two issues were solved \emph{ad hoc}, but we believe the method is readily adaptable to other parameter regions.

An additional computational challenge is scaling the algorithm to
nudge template banks with millions of templates.
The preliminary application of the template nudging algorithm to
the aligned-spin CBC parameter space required the use
of a high-throughput computing cluster. Generating the template isosurface cross-sections in the absence
of a general mismatch metric (analytic or otherwise)
is particularly computationally expensive. The template nudging algorithm is
considerably faster if there is a reliable (and computationally efficient)
way to calculate the metric. By
nudging purely BNS banks (where there is a known analytic expression
of the metric), we were able to nudge an under-saturated stochastic
bank on a single Lenovo Thinkpad T450s Ultrabook laptop. Similarly, the
template banks used for Fermi $\gamma$-ray binary pulsar searches in \cite{Pletsch:2012sh} (which contained tens of millions
of templates) could be constructed by the NCA method in a few hours on one HUAWEI RH1288 v3 Server with two 14 core CPUs since there is an analytic expression for the metric. Therefore, 
it is highly desirable to obtain a computationally efficient way
to calculate the metric in order to apply
the template nudging algorithm to larger CBC parameter spaces, or detectors with improved sensitivities.

The so-called $\theta$ coordinates produce a computationally efficient method for calculating the
numeric approximation of the CBC mismatch metric \cite{Ajith:2012mn,Roy:2017qgg}.
These coordinates are only dependent on the masses and effective
spin and may produce more uniform template isosurfaces, better suited to the template nudging algorithm.
Running the template nudging algorithm with
these flatter coordinates may lower the computational cost
and further reduce the final number of required templates for the CBC template bank.

While our current implementations of the metric agnostic template nudging algorithm are computationally inefficient,
they are still a lot more flexible than the original metric dependent 
NCA method \cite{Fehrmann:2014cpa}.
As long as there is a coordinate system in which cross-sections of individual template isosurfaces can be reduced to a collection of two dimensional ellipsoids, this bootstrap metric construction method is completely generalizable to any current or future CBC template bank parameter space. This makes it a potentially versatile option for building higher dimensional template banks that include the effects of precession, tidal deformation, eccentric orbits, and higher order modes.
\begin{acknowledgments}
We are grateful to Tito Dal Canton, Sebastian Khan and Hirotaka Yuzurihara for valuable comments.
This paper has LIGO document number LIGO-P1700427.
\end{acknowledgments}

\bibliographystyle{apsrev4-1}
\bibliography{optimized-NSBH-AS-CBC-template-bank-paper.bib}

\end{document}